\documentclass[11pt,twoside]{article}

\usepackage{asp2004}
\usepackage{epsfig}
\usepackage{lscape}

\begin{document}
\title{Probing Star Formation in Interacting Galaxies Using UV and Mid-IR: The Case of Arp 82}   
\author{M. Hancock$^1$, B. J. Smith$^1$, C. Struck$^2$, M. Giroux$^1$, P. Appleton$^3$, W. Reach$^3$, V. Charmandaris$^4$}   
\affil{$^1$Department of Physics, Astronomy, and Geology, East Tennessee State University, Box 70652, Johnson City, TN 37614}    
\affil{$^2$Department of Physics and Astronomy, Iowa State University, Ames, IA 50011}
\affil{$^3$Spitzer Science Center, California Institute of Technology, Pasadena, CA 91125}
\affil{$^4$Department of Physics, University of Crete, 71003 Heraklion, Greece}

\begin{abstract} 
To help understand the effects of galaxy interactions on star
formation, we analyze Spitzer infrared and GALEX ultraviolet images
of the interacting galaxy pair Arp 82 (NGC 2535/6), and compare to a
numerical simulation of the interaction.  We investigate the UV and
IR properties of several star forming regions (clumps).  Using the
FUV/NUV colors of the clumps we constrain the ages.  The 8 $\mu$m
and 24 $\mu$m luminosities are used to estimate the far-infrared
luminosities and the star formation rates of the clumps.  We
investigate possible gradients in the UV and IR colors.  See
Smith et al. (2006a,b) for global results on our entire interacting
sample.  
\end{abstract}


\section{Introduction}   

We are investigating whether or not interacting but not yet merging
galaxies have heightened star formation properties.  In our Spitzer
Spirals, Bridges, and Tails interacting galaxy study
(\citealp{smi06a, smi06b}), we have compiled a sample of interacting galaxies
selected from the Arp Atlas of Peculiar Galaxies \citep{arp66}.  We
have previously  presented a detailed study of one of these galaxies,
Arp 107, in  \citet{smi05}.  In the current proceeding we investigate
a second system, the interacting pair Arp 82 (NGC 2535/6)
\citep{han06}.   We have obtained UV, visible, and IR images of Arp 82
from GALEX, SARA, and Spitzer telescopes respectively.

\begin{figure}[ht]
  \begin{center}
    \epsfig{file=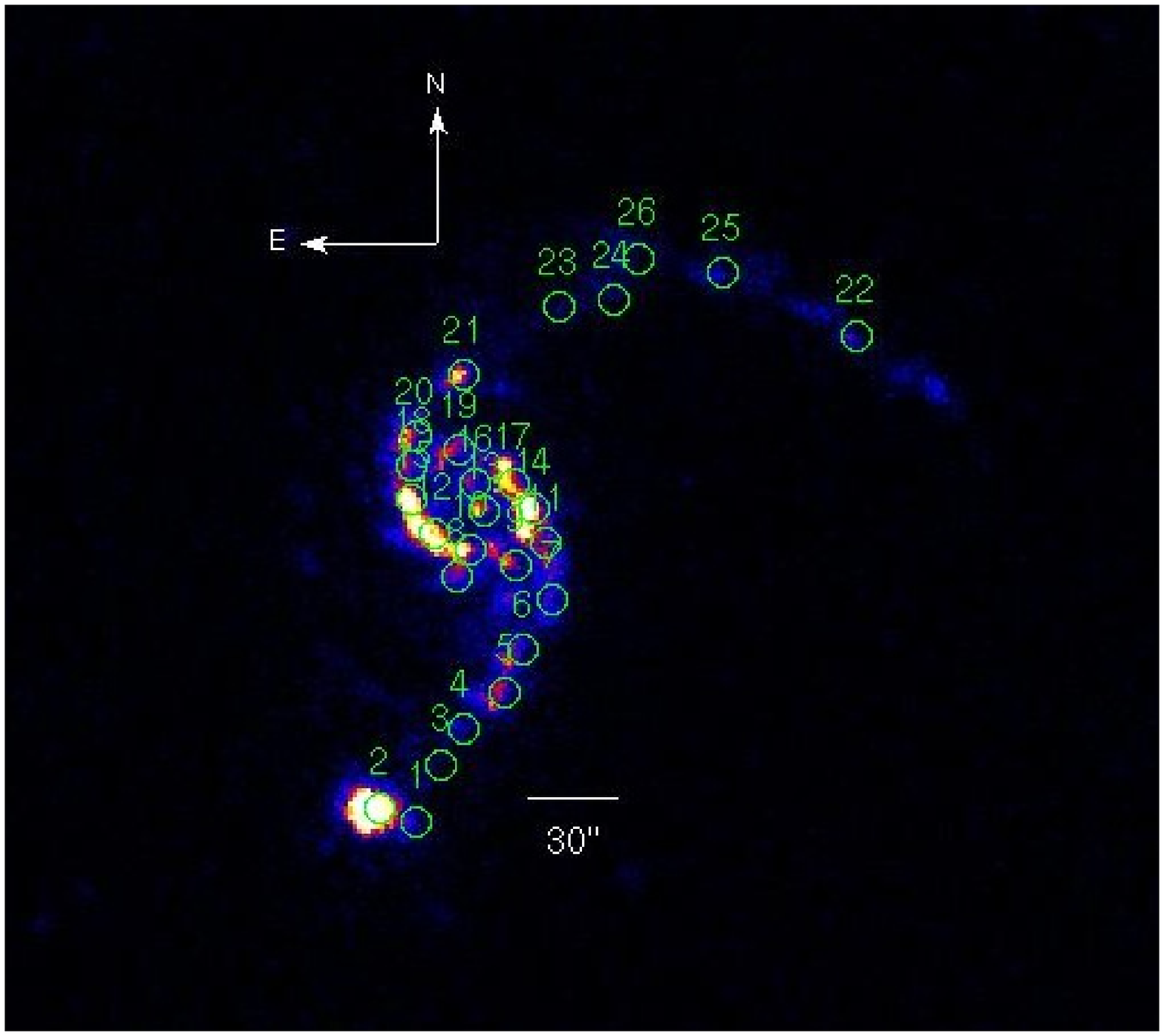, width=5.5cm, height=4.5cm}
    \epsfig{file=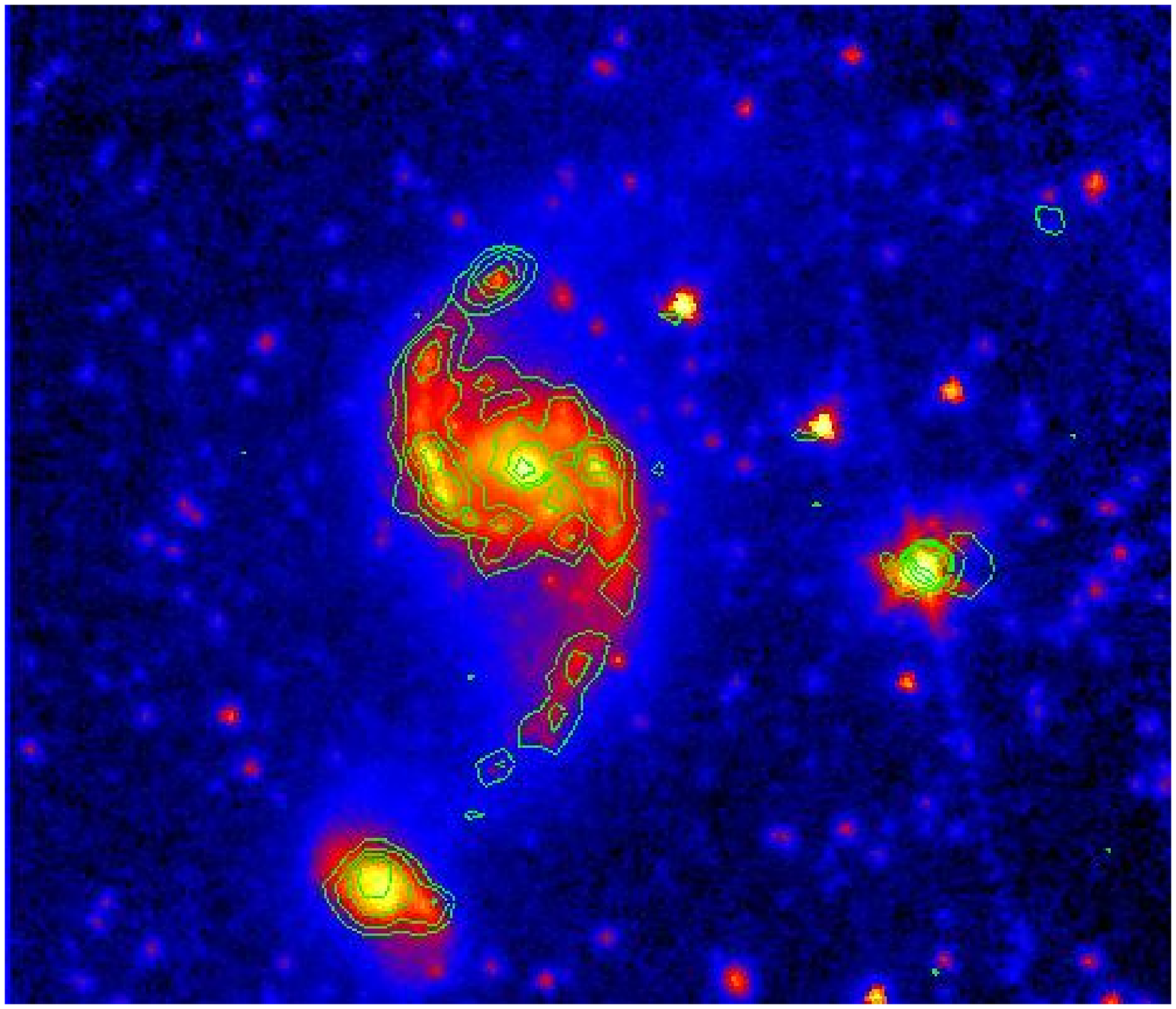, width=5.5cm, height=4.5cm}
    \epsfig{file=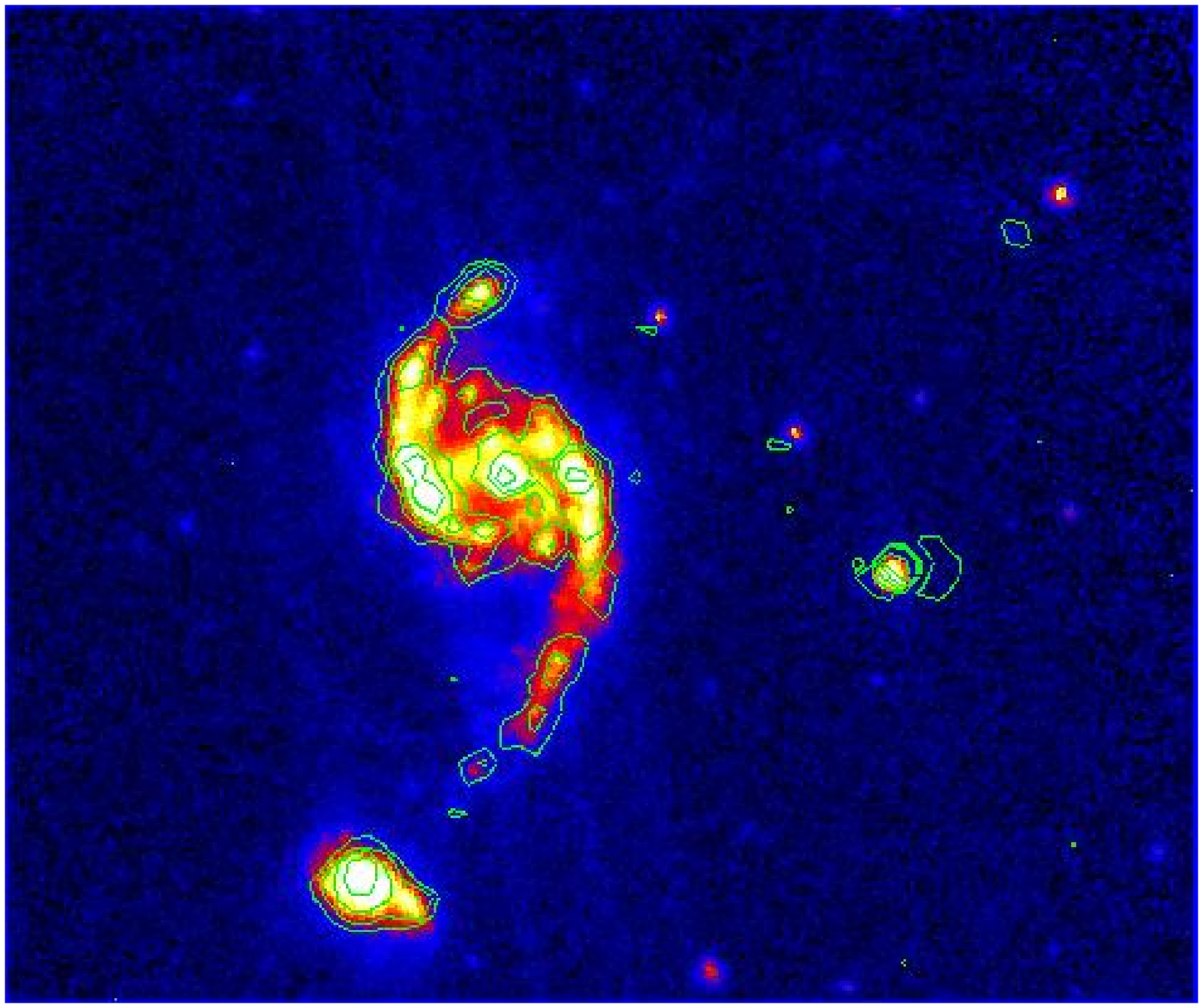, width=5.5cm, height=4.5cm}
    \epsfig{file=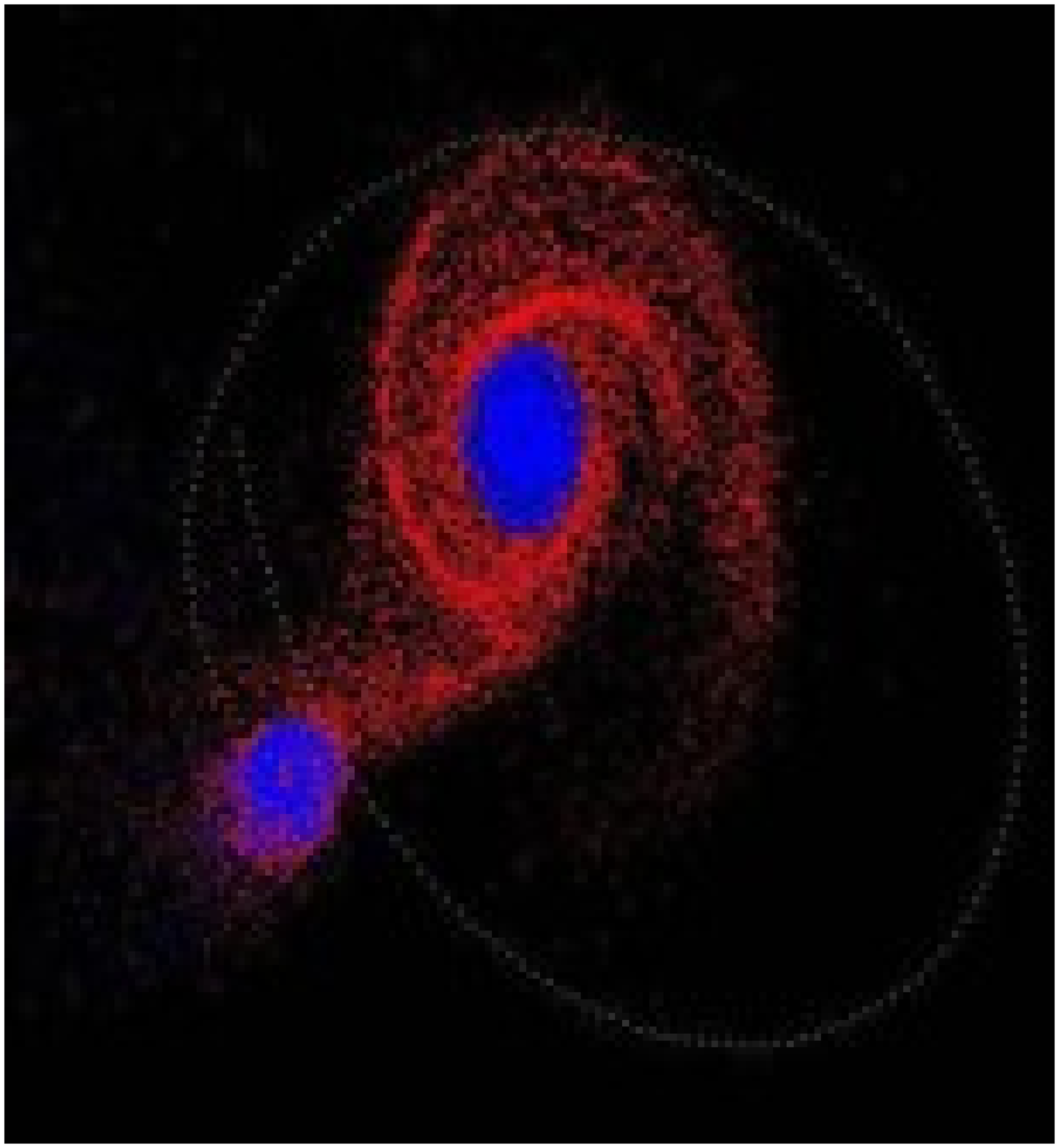, width=5.5cm, height=4.5cm}
  \end{center}
 \vspace{-0.3in}
\caption{From Top (left to right): GALEX FUV image, Spitzer 3.6 $\mu$m image with SARA H$\alpha$ contours, Spitzer 8.0 $\mu$m image with SARA H$\alpha$ contours, and a smooth particle hydrodynamics model (see text).  North is up and east is to the left.  At the distance of Arp 82 (57 Mpc) 30\arcsec\ corresponds to $\sim$8.3 kpc.}
\end{figure}

\begin{figure}[ht]
  \begin{center}
    \epsfig{file=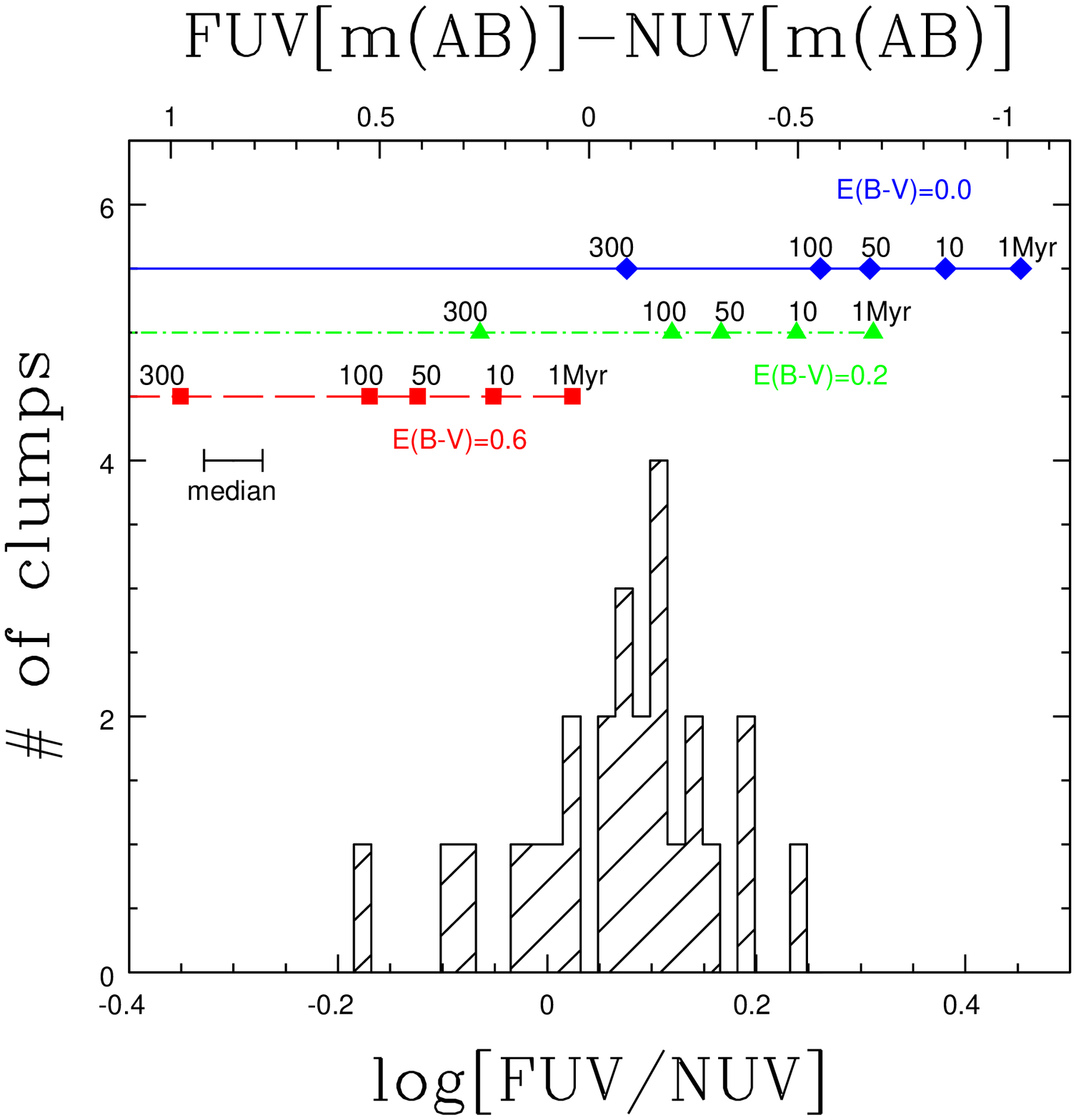, width=5.5cm}
    \epsfig{file=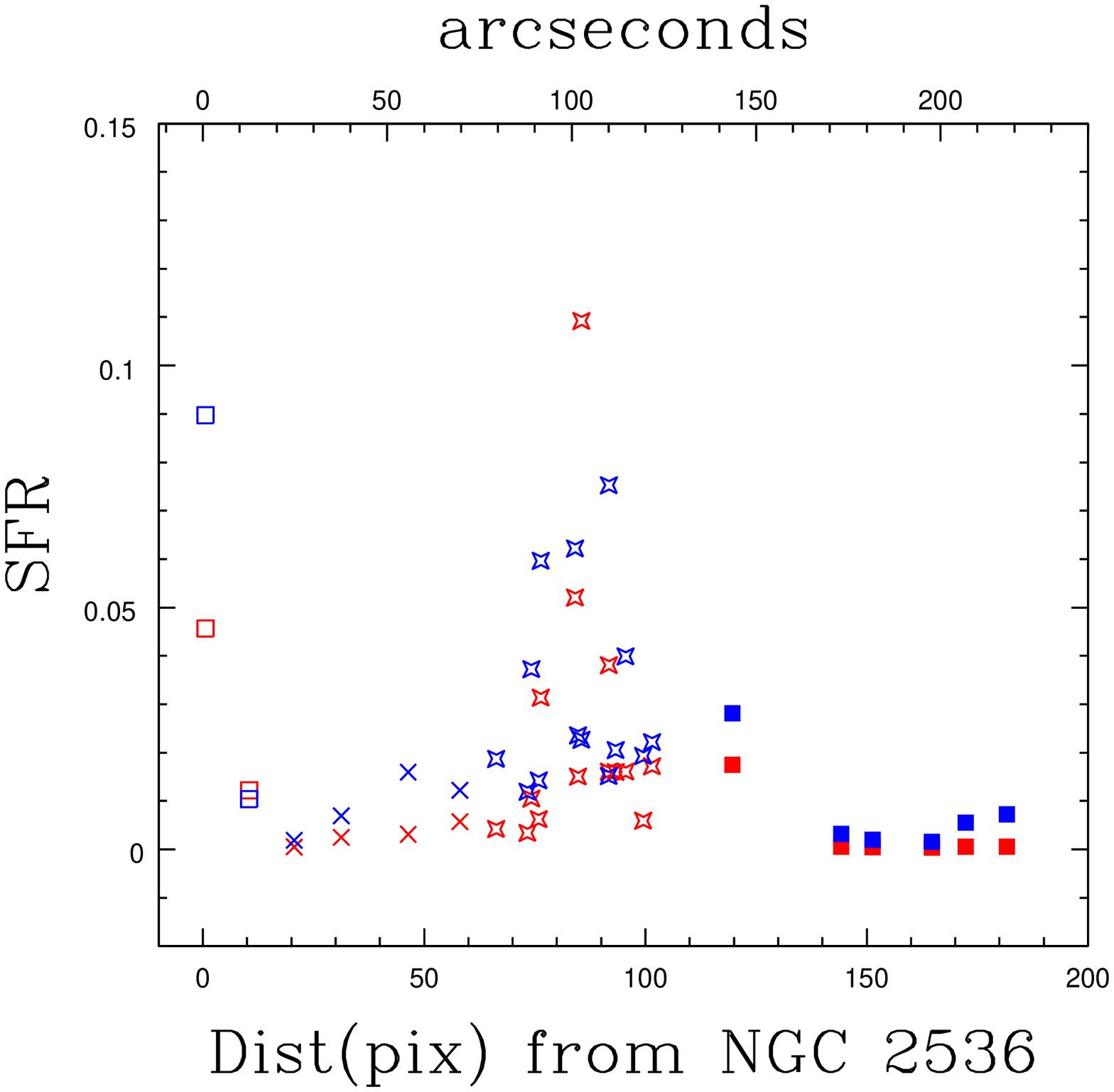, width=5.5cm}
    \epsfig{file=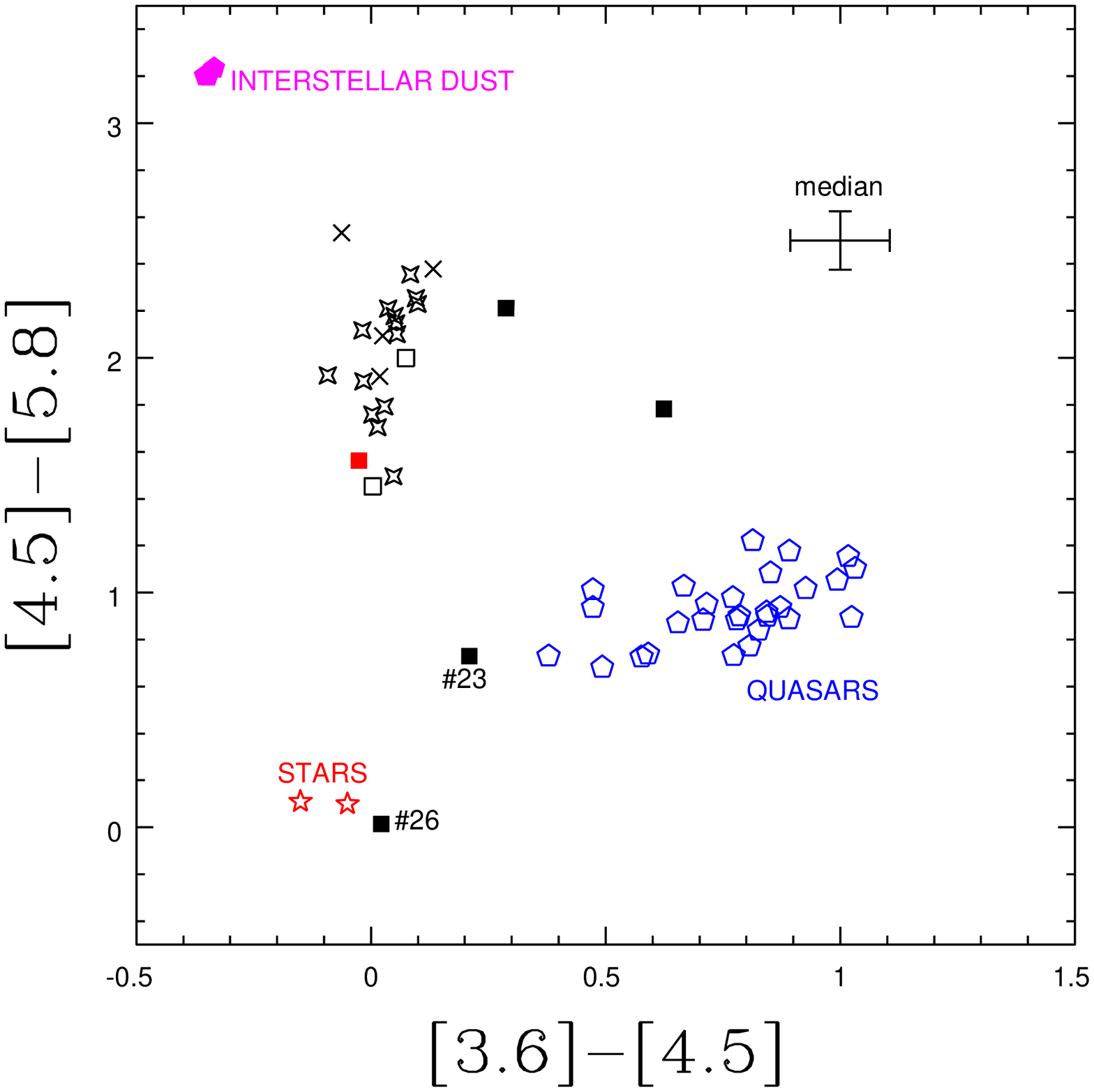, width=5.5cm}
    \epsfig{file=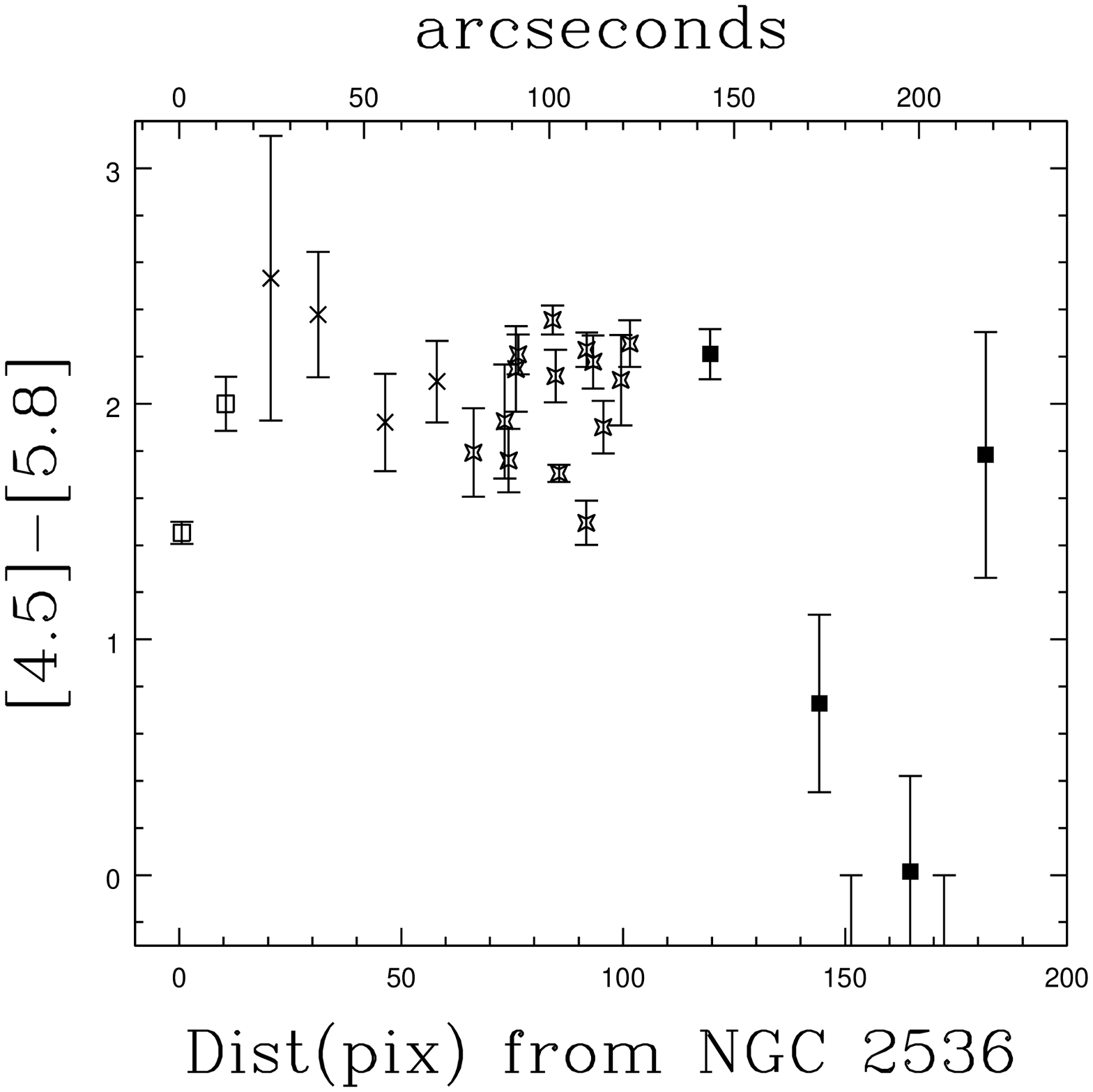, width=5.5cm}
  \end{center}
 \vspace{-0.3in}
\caption{From Top (left to right): FUV/NUV distribution, SFR vs Distance, [4.5]-[5.8] vs [3.6]-[3.4], [4.5]-[5.] vs distance (see text).}
\end{figure}

\section{Discussion}

Figure 1 displays various images of Arp 82.  The top left is a GALEX
far-UV image with the 26 clumps identified.   The northern galaxy is
NGC 2535 and southern galaxy is NGC 2536.  The top right image in
Figure 1 is Arp 82 in the Spitzer IRAC 3.6 $\mu$m band with H$\alpha$
contours from the SARA telescope.  The bottom left image in Figure 1
is Arp 82 in the Spitzer IRAC 8 $\mu$m  band with SARA H$\alpha$
contours.  The tail is more prominent in the UV than in the IR while
the center is much less prominent.    Note  that there are FUV and
8 $\mu$m clumps in the tail region that are  not seen in H$\alpha$.
The star forming regions at 8 $\mu$m and in the FUV are more prominent
than at 3.6 $\mu$m.

The bottom right image in Figure 1 is a snapshot of a smooth particle
hydrodynamics model of the gas in red and old stars in blue.   This
image shows Arp 82 about 1 Gyr  after the initial closest approach.
The dotted curve shows the companion's passage.  The orbit is nearly
planar.  The long duration is needed to allow particles to propagate
out to the large distances observed.

Four individual plots are seen in Figure 2.  The top left plots the
GALEX FUV/NUV distribution.  A Starburst99 \citep{lei99} stellar
population synthesis model reddened with E(B-V)=0.0(blue), 0.2(green),
and 0.6(red) mag according to the \citet{cal94} reddening law is
shown at the top of the histogram.  Selected ages are marked.  The top
axis is in magnitudes.  Most of the clumps have an E(B-V) between 0.2
and 0.6 mag and ages $<300$ Myr while a few clumps may be $<10$ Myr.

We have determined the star formation rate (SFR) for the clumps using
two independent methods.  First, we estimated them from the L(IR)
using the calibration in \citet{ken98}.  The clumps have a total
SFR$_{IR}$ of $\sim0.43$ M$_\odot$ yr$^{-1}$.   Second, we estimated
the SFR from the L$_{\nu}$(FUV) using the  UV SFR calibration in
\citet{ken98}.  The total SFR$_{UV}$ of the clumps is $\sim0.63$
M$_\odot$ yr$^{-1}$,  in good agreement with the SFR$_{IR}$.  No reddening
correction has been applied to the L$_{\nu}$(FUV).

The top right plot in Figure 2 shows the star formation rate (SFR) versus
distance from NGC 2536.  The red symbols are the SFR determined from
L(IR) and the blue  symbols are the SFR determined from
L$_{\nu}$(FUV).  The open boxes represent clumps in NGC 2536, x's
represent clumps in the bridge region, stars represent clumps in the
spiral (NGC 2535) region, and filled boxes represent clumps in the tail
region.  The top axis is in arcseconds.  

From this figure it can be seen that  the SFR is greatest in the
spiral region of NGC 2535 and in NGC 2536,  with much less star
formation in the bridge and  tail regions.  It can
also be seen that the SFR's of clumps in the bridge and tail regions
have much better agreement than do the SFR's of clumps in NGC 2536 and
the spiral regions.  In most cases the SFR$_{UV}$ is greater than the
SFR$_{IR}$.  If an extinction correction were applied to the
L$_{\nu}$(FUV) the SFR$_{UV}$ would be greater and the agreement with
SFR$_{IR}$ would be worse.

The clumps in the bridge and tail regions account for about 7\% of the
total clump SFR$_{IR}$, while the 2 clumps in the small companion, NGC
2536, and the 2 largest clumps in the spiral region of NGC 2535 (\#13
and \#16) make up about 42\% of the total clump SFR$_{IR}$.  The
SFR$_{IR}$ of the entire Arp 82 system is 1.2 M$_\odot$ yr$^{-1}$,
while the entire system SFR$_{UV}$ is 2.7 M$_\odot$ yr$^{-1}$.   The
total clump SFR$_{IR}$ accounts for about 36\% of the entire system
SFR$_{IR}$.

The bottom left graph in Figure 2 plots the IRAC [4.5]$-$[5.8] vs
[3.6]$-$[4.5] colors of the clumps.  The data symbols are the same as
above.  Also included in this figure are the
predicted IRAC colors for interstellar dust \citep{li01}, the Sloan
Digitized Sky Survey quasars in the Spitzer Wide-Area Infrared
Extragalactic Survey (SWIRE) Elais N1 field \citep{hat05}, and the
colors of M III stars from M. Cohen (2005, private communication) and
field stars from \citet{whit04}.  The quasars have redshifts between
0.5 and 3.65; since their spectral energy distributions are power
laws, their infrared colors do not vary much with redshift.  From
these figures, it can be seen that clumps \#23 and \#26 have colors
consistent with those of quasars and field stars respectively and may
not be part of Arp 82.

Most of the clumps have [4.5]$-$[5.8] colors between those of the ISM
and stars, indicating contributions from both to this color.  Clumps
\#24 and \#25, which are in the northern tail, have colors similar to
those of ISM (but with large uncertainties). Thus these appear to be
very young star formation regions with little underlying old stellar
population.

The bottom right graph in Figure 2 plots the IRAC [4.5]$-$[5.8] color vs
distance from NGC 2536.  The data symbols and horizontal axis are the
same as above.  From this plot it can be seen that clumps in the
bridge  and tail regions seem to have different relative ages than
those in the  spiral region.  The [4.5]$-$[5.8] colors are generally
very red (i.e., very `starbursty'),  except for the two low S/N clumps
in the tail, \#23 and \#26.

\acknowledgements 

This work is based in part on observations made with the Spitzer
Space Telescope, which is operated by the Jet Propulsion Laboratory,
California Institute of Technology under contract with NASA. 
GALEX is a NASA Small Explorer mission, developed in cooperation with the
Centre National d'Etudes Spatiales of France and the Korean Ministry of 
Science and Technology.  This research was supported by NASA Spitzer grant
1263924, NSF grant AST 00-97616, NASA LTSA grant NAG5-13079, and GALEX
grant GALEXGI04-0000-0026.  This work has made use of the
NASA/IPAC Extragalactic Database (NED), which is operated by the Jet
Propulsion Laboratory, California Institute of Technology, under
contract with NASA.


\end{document}